\documentclass[letterpaper,twocolumn,10pt]{article}
\usepackage{usenix}
\usepackage{CJKutf8}

\usepackage{tikz}
\usepackage{amsmath}
\usepackage{graphicx}
\usepackage{filecontents}

\usepackage{wasysym}
\usepackage[table]{xcolor}
\usepackage{algorithm}    
\usepackage{algorithmic}  
\usepackage{xcolor}       

\begin{document}
\begin{CJK}{UTF8}{gbsn}

\date{}

\title{\Large \bf MirageNet:A Secure, Efficient, and Scalable On-Device Model Protection in  Heterogeneous TEE and GPU System}

\author{
{\rm Huadi Zheng\footnotemark[2]}\\
National University of Defense Technology
\and
{\rm Cheng Li\footnotemark[2]}\\
National University of Defense Technology
\and
{\rm Yan Ding\footnotemark[1]}\\
National University of Defense Technology
} 

\maketitle

\begin{abstract}
With the growing computational capabilities of edge devices, deploying high-performance deep neural network (DNN) models on untrusted third-party hardware has become a practical approach to reduce inference latency and safeguard user data privacy. Given the high cost of model training and the need to preserve user experience, achieving model privacy protection with low runtime overhead is a key challenge. Leveraging Trusted Execution Environments (TEEs) offers a pragmatic defense path. To this end, prior work has proposed heterogeneous GPU–TEE inference frameworks based on parameter obfuscation to balance GPU compute efficiency and model confidentiality. However, recent studies reveal that several partial obfuscation defenses have become ineffective, while existing robust schemes introduce unacceptable additional latency.

To address these challenges, we propose ConvShatter, a novel model obfuscation scheme that achieves low inference latency and high accuracy while preserving model confidentiality and integrity. ConvShatter exploits the linearity of convolution to decompose each original kernel into critical kernels and common kernels, and injects confounding decoy kernels. In addition, it permutes the channel order within each kernel and the kernel order within each convolutional layer. Prior to deployment, ConvShatter performs kernel decomposition, decoy injection, and order obfuscation, and securely stores the minimal recovery parameters in the TEE. During inference, the TEE reconstructs the outputs of each obfuscated convolutional layer.

Extensive experiments demonstrate that ConvShatter substantially reduces inference latency overhead while maintaining strong confidentiality and integrity guarantees. Compared with existing schemes offering comparable protection, ConvShatter reduces overhead by 16\% (replace with the concrete value or percentage), while preserving inference accuracy on par with the original model.

\begingroup
\renewcommand{\thefootnote}{\textdagger}
\footnotetext{Cheng li and Huadi Zheng contribute equally. }
\endgroup
\end{abstract}

\section{Introduction}

\begin{figure}[t]
  \centering
  \includegraphics[width=\linewidth, keepaspectratio]{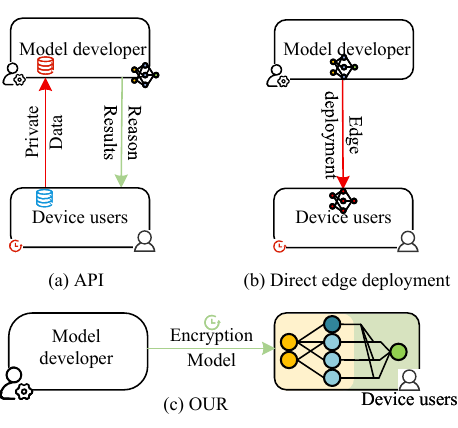}
  \caption{Illustration of typical interaction modes between users and models. (a) API: Users interact with models via API endpoints. (b) Edge deployment: Developers deploy models directly on the user side for inference. (c) Our approach: Models are encrypted before deployment, with decryption keys securely stored in the user's trusted execution environment (TEE).}
  \label{fig:guanxi}
\end{figure}

With the increasing adoption of deep learning in fields such as document recognition and computer vision~\cite{13,14,15,16}, the value of models has become increasingly prominent. Cloud deployment can lead to significant latency during user interaction, prompting developers to prefer model deployment on edge devices for local inference. Researchers have also proposed various approaches for accelerating edge computing~\cite{17,18,19}. However, edge deployment introduces critical security and privacy concerns~\cite{1}; the user side is inherently untrusted, and unprocessed models deployed on the edge become white-box, exposing all model information to potential adversaries masquerading as users~\cite{2,6}. 

Figure~\ref{fig:guanxi}(a) illustrates cloud-based deployment, where user data must be transmitted to the cloud for inference. While the model is fully protected, users face data privacy risks and high interaction latency~\cite{10,11}, resulting in compromised user experience. Figure~\ref{fig:guanxi}(b) shows direct edge deployment with no protection, rendering model parameters fully transparent to users and resulting in unacceptable leakage of valuable intellectual property~\cite{12}. Figure~\ref{fig:guanxi}(c) presents our approach: encrypted models are deployed on the edge, with decryption keys confined to the secure TEE. Only authorized users can access the keys and decrypt the model, preventing direct theft~\cite{MLCapsule}. 

To enable this paradigm, prior works have explored heterogeneous GPU-based TEEs~\cite{Heterogeneous,Graviton,GOAT} and GPU-TEE system designs~\cite{Goten}. However, heterogeneous GPU designs suffer from plaintext model exposure~\cite{GROUPCOVER} and practical deployment challenges~\cite{SOTER}. Thus, GPU-TEE based systems have become the research focus, spawning numerous variants~\cite{TEESlice,ShadowNet,DarkneTZ,SOTER,NNSplitter,eNNclave,GROUPCOVER,Slalom,Magnitude}, where TEE stores decryption secrets or masks, and most linear computations are offloaded to the GPU. Due to limited computation and storage in the TEE, minimizing TEE-resident data for secure protection is a key challenge.

The prevailing solution encrypts model weights via obfuscation~\cite{TEESlice,ShadowNet,DarkneTZ,SOTER,NNSplitter,eNNclave,GROUPCOVER,Slalom,Magnitude}: the edge user only receives obfuscated weights, while decryption routines are stored in the TEE, accessible only with developer authorization. Adversaries, even with full GPU-side visibility, cannot directly infer correct outputs and must resort to black-box model extraction attacks~\cite{3,4,5,knockoff}, raising the cost and complexity of model theft and enhancing model privacy.

Existing obfuscation methods fall into full-weight obfuscation and partial obfuscation schemes. Partial obfuscation targets only a subset of weights, reducing overhead but relying on the secrecy of the obfuscated subset~\cite{NNSplitter}. For example, NNSplitter~\cite{NNSplitter} obfuscates merely 0.002\% of weights, leaving most weights exposed. If the adversary can identify the obfuscated weights, a near-perfect replica can be reconstructed, rendering partial obfuscation ineffective. Recent work, groupcover~\cite{GROUPCOVER}, reveals that sorting trained weights produces a smooth curve, but obfuscating important weights disrupts this smoothness, creating clusters that betray the obfuscated region. By isolating this region, adversaries can retrain only the obfuscated weights, achieving highly effective model stealing against partial schemes.

Full-weight obfuscation encrypts all private weights, with decryption secrets stored in the TEE, enabling layer-by-layer recovery of intermediate results. While obfuscated weights are non-transparent, full obfuscation incurs substantial computational overhead. Moreover, recent research~\cite{GROUPCOVER} demonstrates that high cosine similarity persists between pre-trained and post-trained weights, enabling attacks that reverse simple linear transformations, as in Soter~\cite{SOTER}. Groupcover further improves secrecy by leveraging mutual coverage among weights, but the high overhead remains a challenge for practical deployment.

The crux is balancing weight confidentiality and minimal obfuscation overhead. To address this, we propose ConvShatter, a convolutional kernel decomposition-based obfuscation scheme, which effectively mitigates the cosine similarity vulnerability with low extra cost.

Our contributions are summarized as follows:

\begin{itemize}
    \item We propose a novel obfuscation method based on convolutional kernel decomposition, the first of its kind, enabling secure encryption of model weights in untrusted environments.
    \item We implement a complete edge deployment inference system, where the obfuscated model efficiently runs in GPU-TEE systems, with almost all computation offloaded to the GPU and minimal TEE overhead.
    \item Against state-of-the-art model extraction attacks, we introduce intra-kernel channel confusion, which prevents adversaries from identifying and recovering obfuscated weights, and demonstrate via extensive experiments that SOTA model stealing accuracy is reduced to random guessing with negligible overhead.
\end{itemize}

\section{Background}

\subsection{Model Stealing}

This work focuses on defending against two adversarial parties: model thieves and victim models. In model stealing attacks, adversaries are assumed to have full access to public model architectures~\cite{7} and all deployment information outside the secure environment. Through membership inference~\cite{8,9} and other techniques, adversaries may also obtain part of the original training data. In the absence of protection, the model is fully exposed, enabling direct theft. After protection, the attack reduces to a black-box scenario, where adversaries query the model via API to collect input-output pairs and train a replica model~\cite{4}. In this work, we leverage groupcover~\cite{GROUPCOVER}'s weight clustering and weight cosine similarity to assist adversaries in identifying obfuscated weights, enabling optimized extraction. This approach is especially effective against partial obfuscation, where most weights remain unprotected, and identifying obfuscated weights allows the adversary to recover a high-fidelity clone.

Groupcover~\cite{GROUPCOVER} shows that sorting weights reveals smooth curves, but partial obfuscation introduces anomalies, making obfuscated weights identifiable. In full obfuscation, high cosine similarity between public and private weights allows adversaries to locate obfuscated weights and recover them via simple linear transformations. Combining these with knockoff~\cite{knockoff} model stealing attacks yields near-identical clone models.

\subsection{Trusted Execution Environment (TEE)}

TEE is a trusted secure enclave. With the rise of edge deployment, TEE technology has matured~\cite{Ding,OP-TEE,Song}. In our experiments, the TEE is assumed to be perfectly secure; adversaries cannot access any information within, while everything outside is fully visible~\cite{No}. Due to limited TEE storage and computation, only minimal sensitive data and computation should reside within. Each time decryption is required, data must be transferred into the TEE for computation, incurring significant overhead. Therefore, minimizing TEE use is key to reducing obfuscation overhead while maintaining confidentiality.

\subsection{Victim Models}

Victim models are the targets of model theft. Due to excessive latency in cloud deployment, victim models are deployed at the edge, increasing vulnerability. Thus, protection via TEE is necessary. Adversaries can still access the public model architecture and exposed weights; hence, encryption is required for the externally deployed model. Victim model defense relies on protecting the decryption keys within the TEE~\cite{20}. The victim's goal is to prevent low-cost extraction of high-value model information, while most data must remain outside the TEE for performance reasons. In full obfuscation, all weights are encrypted; in partial obfuscation, many weights are unprotected, but adversaries do not know which. Finding obfuscated weights is crucial for successful extraction.

Adversary's goal: extract model information to replicate a functionally equivalent clone. Adversaries are assumed to observe a subset of input-output pairs, access all public model data and external weights, and know the protection scheme. Adversaries may first attempt de-obfuscation and then perform knockoff stealing. For comparison, we also evaluate cases with limited access (e.g., 1\% of training data~\cite{tensorshield}) to validate effectiveness, noting that adversaries cannot obtain private user input.

\subsection{Adversarial Capabilities}

We follow established threat models~\cite{knockoff}, assuming the adversary has maximum capability: access to 10\% of the original training data and all information outside the TEE. The adversary knows the victim's model architecture, accesses corresponding public models, and is aware of the protection scheme, enabling targeted de-obfuscation prior to knockoff extraction. We conduct comparative experiments with lower adversary access (e.g., 1\%) to demonstrate effectiveness. As the model is deployed at the edge, adversaries cannot access personal user input.

\subsection{Existing Obfuscation Schemes}

Partial obfuscation approaches such as Magnitude~\cite{Magnitude} and NNSplitter~\cite{NNSplitter} target the most ``important'' weights. NNSplitter uses reinforcement learning to analyze weight importance and replaces selected weights with random values in a specified interval, while Magnitude simply selects the largest-magnitude weights for replacement. DarkneTZ~\cite{DarkneTZ} partitions the model, placing sensitive layers entirely within the TEE. Full obfuscation methods include eNNclave~\cite{eNNclave}, which trains a few layers and runs them in the TEE, while untrained public layers run on the GPU. Slalom~\cite{Slalom} uses masking to obfuscate intermediate results. Soter~\cite{SOTER}, GroupCover~\cite{GROUPCOVER}, and ShadowNet~\cite{ShadowNet} further enhance security: Soter introduces blinding via multiplicative masks, GroupCover applies obfuscation matrices and masking, and ShadowNet adds noise and scaling to all weights, which is removed in the TEE. Existing partial obfuscation schemes have proven vulnerable to identification of obfuscated weights, while full obfuscation schemes generally suffer from high overhead; only GroupCover achieves a better security-overhead balance.

\begin{table}[ht]
  \centering
  \caption{Comparison of existing edge security deployment schemes}
  \label{tab:comparison}
  \renewcommand{\arraystretch}{1.2}
  \resizebox{0.8\columnwidth}{!}{%
  \begin{tabular}{lccccc}
  \hline
  Project    & Security & Accuracy & Pre-cost & Inference-cost & Overall \\ 
  \hline
  Darknight  & High & Medium   & Medium   & Medium  & Low \\
  eNNclave   & High & Medium   & Medium   & Medium  & Low \\
  Magnitude  & Low  & Low      & Low      & Medium  & Low \\
  NNSplitter & Low  & High     & Low      & Medium  & Low \\
  Soter      & Medium & Low    & Medium   & Medium  & Medium \\
  Slalom     & Low  & High     & High     & Medium  & Low \\
  GroupCover & Low  & Medium   & High     & Medium  & Low \\
  Ours       & High & High     & Low      & Medium  & High \\ 
  \hline
  \end{tabular}%
  }
\end{table}

Table~\ref{tab:comparison} provides a systematic evaluation of these obfuscation schemes.


\section{System Overview}

\begin{figure}[ht]
  \centering
  \includegraphics[width=\linewidth, keepaspectratio]{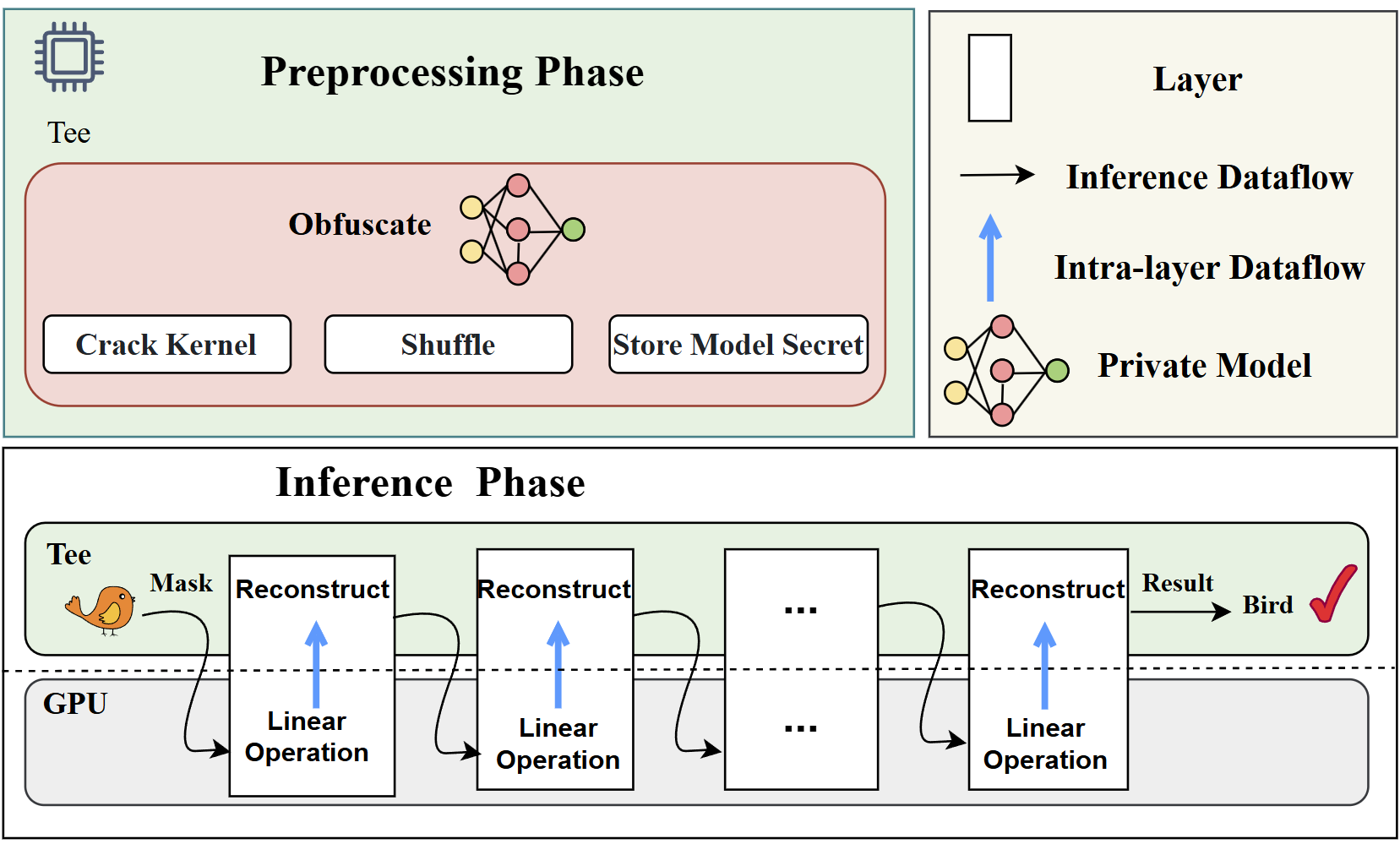}
  \caption{Overall workflow of the proposed obfuscation framework.}
  \label{fig:framework}
\end{figure}

\begin{figure*}[ht]
  \centering
  \includegraphics[width=\linewidth]{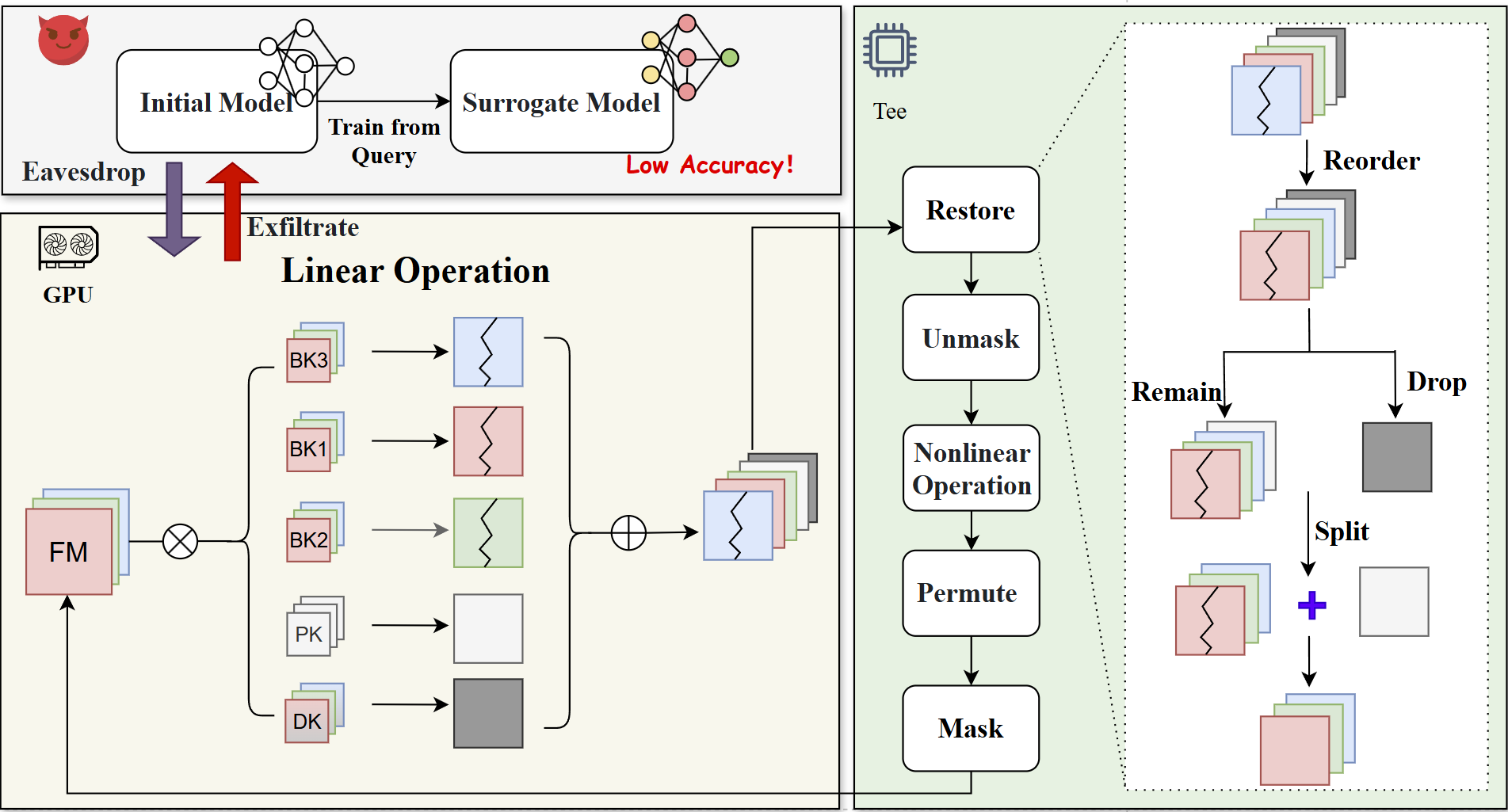}
  \caption{Inference process of the proposed obfuscation framework.}
  \label{fig:detail_inference}
\end{figure*}

In this section, we present ConvShatter, a model protection mechanism tailored for TEE–GPU heterogeneous inference that perturbs both the structural and statistical regularities of convolutional kernels in pre-trained and private models. As summarized in Figure~\ref{fig:framework}, the framework comprises an offline obfuscation phase and an online secure inference phase designed to maximize GPU utilization while minimizing TEE residency and communication.

\subsection{Design Goals}
ConvShatter targets four primary goals:
- Confidentiality: Prevent disclosure of proprietary weights and layer topology from the untrusted GPU and host.
- Integrity: Guarantee that the recovered forward pass in the TEE matches the baseline model semantics up to numerical tolerance.
- Efficiency: Bound added latency and memory overhead to be acceptable for real-time edge workloads.
- Composability: Integrate with standard deployment stacks (e.g., cuDNN/TensorRT) with minimal engineering friction.

\subsection{Threat Model and Security Guarantees}
We assume a powerful adversary who:
- Observes all data and control paths outside the TEE, including GPU-visible weights, activations, and intermediate results.
- Knows the architecture family, public pre-trained counterparts, and the obfuscation algorithmic blueprint (Kerckhoffs’ principle).
- May obtain up to 10\% of training data and mount black-box queries to train a surrogate (model extraction).

The adversary cannot:
- Read TEE state (keys, permutations, recovery coefficients).
- Tamper with TEE execution integrity.

Security objective: the leakage function L over the attacker’s view should be information-theoretically or computationally insufficient to reconstruct (i) kernel alignments (channel and kernel permutations), (ii) kernel identities (distinguishing patch vs. decoy), or (iii) deobfuscation coefficients with accuracy sufficient to recover baseline accuracy beyond a black-box surrogate. Our defense aims to render similarity-based alignment statistically indistinguishable and increase the sample complexity of deobfuscation beyond practical limits.

\subsection{Offline Obfuscation Phase}
\begin{algorithm}[t]
\caption{Weight Obfuscation with Enhanced Fake Kernel Obfuscation}
\label{alg:weight_obfuscation}
{\raggedright
\textbf{Input}: Pre-trained weights $\mathbf{W} = \{W_1, W_2, \dots, W_L\}$ for $L$ layers; Number of public kernels $k_{pub}$; Number of fake kernels $k_{fake}$

\textbf{Output}: Obfuscated weights $\mathbf{W}^{obf}$, Precomputed noise $\mathbf{M}$
\par}
\begin{algorithmic}[1]
\STATE Generate a global random mask $m$ for the input domain
\FOR{each convolutional layer $l$ to be obfuscated}
    \STATE Generate a random channel permutation key $\pi_l$ for the input channels
    \STATE $\mathbf{W}_l \gets \text{shuffle\_channels}(\mathbf{W}_l, \pi_l)$
    \STATE Initialize obfuscated weight set: $\mathbf{W}_l^{obf} \gets \emptyset$
    \STATE Set $k_{total} = k_{pub} + k_{fake}$
    \STATE Generate $k_{total}$ random kernels $\mathbf{K} = [K_1, K_2, \dots, K_{k_{total}}]$ with the same random initialization method
    \STATE Randomly shuffle the indices of $\mathbf{K}$ to form a permutation $\tau$
    \STATE Generate a shared random coefficient vector: $\mathbf{c} \gets [c_1, c_2, ..., c_{k_{pub}}] \sim \mathcal{U}(0,1)$
    \FOR{each original kernel $W_l^i$ in layer $l$}
        \STATE Compute kernel-specific component: $W_l^{i'} \gets W_l^i - \sum_{j=1}^{k_{pub}} c_j K_{\tau(j)}$
        \STATE Add $W_l^{i'}$ to $\mathbf{W}_l^{obf}$
    \ENDFOR
    \FOR{$j=1$ to $k_{pub} + k_{fake}$}
        \STATE Add $K_{\tau(j)}$ to $\mathbf{W}_l^{obf}$
    \ENDFOR
    \STATE Precompute the noise: $\mathbf{M}_l \gets \text{Conv}(\mathbf{W}_l^{obf}, m)$
\ENDFOR
\STATE Securely save coefficients $\mathbf{c}$, channel permutation keys $\{\pi_l\}$, and precomputed noise $\mathbf{M}$ for TEE
\RETURN $\mathbf{W}^{obf}$, $\mathbf{M}$
\end{algorithmic}
\end{algorithm}
During offline obfuscation, each convolutional layer is reparameterized by shared patch bases, per-output damaged components, and injected decoys:

\begin{equation}
  W_i = B_i + \sum_{k=1}^{K} \alpha_{i,k} P_k,
\end{equation}
where $W_i$ denotes the $i$-th kernel, $B_i$ the damaged kernel, $P_k$ a shared patch basis, and $\alpha_{i,k}$ the coefficients. We construct $\{P_k\}$ by QR/Gram–Schmidt orthogonalization applied to linear combinations of original kernels with randomized mixing, followed by scale-normalization to match per-filter statistics. Decoys are sampled from the span of $\{P_k\}$ with randomized coefficients and spectral shaping to match empirical distributions (norms and kurtosis) of the original filters.

To suppress position-wise similarity leakage, we apply:
- Channel permutations within kernels and kernel-order permutations within the layer.
- Spectrum-preserving scaling on $B_i$ to decorrelate with any public pre-trained kernels while retaining reconstructability via TEE-side compensation.

Recovery metadata (permutations, seed material for decoy generation, low-dimensional coefficients, and per-layer digests) is sealed and provisioned to the TEE. All GPU-visible tensors are obfuscated and statistically camouflaged.

\subsection{Secure Inference Phase}
\begin{algorithm}[t]
\caption{Secure Inference with Full Kernel Confusion (Additive Input Masking)}
\label{alg:secure_inference_gpu_all_kernels_add}
\textbf{Require}: Input data $\mathbf{x}$; Obfuscated weights $\mathbf{W}^{obf}$; Precomputed noise $\mathbf{M}$; Shared coefficients $\mathbf{c}$; Channel permutation keys $\{\pi_l\}$; Number of public kernels $k_{pub}$; Number of fake kernels $k_{fake}$; Output mask $\mathbf{m}_{out}$; Kernel index permutation $\tau$\\
\textbf{Ensure}: Final output $\mathbf{y}_{final}$
\begin{algorithmic}[1]
\STATE \textbf{BEGIN}
\FOR{each input $\mathbf{x}$ (or mini-batch)}
    \STATE \textbf{In TEE:}
    \STATE Generate random input noise $\mathbf{m}$
    \STATE Compute masked input: $\hat{\mathbf{x}} = \mathbf{x} + \mathbf{m}$
    \STATE Send $\hat{\mathbf{x}}$ to GPU

    \STATE \textbf{On GPU:}
    \STATE Compute all convolutions:
        \begin{itemize}
            \item $\mathbf{C}_{aux} = \{\text{Conv}(\hat{\mathbf{x}}, K_{\tau(j)})\}_{j=1}^{k_{pub}+k_{fake}}$
            \item $\mathbf{C}_{spec} = \{\text{Conv}(\hat{\mathbf{x}}, W_l^{i'})\}_{i}$
        \end{itemize}
    \STATE Send $\mathbf{C}_{aux}$, $\mathbf{C}_{spec}$ (and $\mathbf{M}$ if needed) to TEE

    \STATE \textbf{In TEE:}
    \STATE Compute public kernel combination: $\mathbf{p} = \sum_{j=1}^{k_{pub}} c_j \cdot \mathbf{C}_{aux}[j]$
    \STATE Initialize reconstructed feature map $\mathbf{f} = \emptyset$
    \FOR{each kernel-specific component $W_l^{i'}$}
        \STATE Reconstruct channel: $f_i = \mathbf{C}_{spec}[i] + \mathbf{p} - \mathbf{M}[i]$
        \STATE Add $f_i$ to $\mathbf{f}$
    \ENDFOR
    \STATE Restore channel order: $\mathbf{y} = \pi_l^{-1}(\mathbf{f})$
    \STATE Apply output mask: $\mathbf{y}_{final} = \mathbf{y} \odot \mathbf{m}_{out}$
    \RETURN $\mathbf{y}_{final}$
\ENDFOR
\STATE \textbf{END}
\end{algorithmic}
\end{algorithm}
At inference time, we minimize TEE involvement by exploiting convolution linearity. Let $\mathbf{X}$ denote the input activation tensor. We compute on the GPU:
- Shared patch feature maps $z_k^{\mathrm{patch}} = \sum_j P_{k,j} * x_j$.
- Damaged components $z_i^{\mathrm{damaged}} = \sum_j B_{i,j} * x_j$.

The TEE then:
- Recombines $z_k^{\mathrm{patch}}$ via the sealed coefficients to form $\sum_k \alpha_{i,k} z_k^{\mathrm{patch}}$.
- Restores channel and kernel orders using the sealed permutations.
- Applies additive one-time pad (OTP) masking on selected intermediates to thwart side-channel or correlation-based reconstruction. Precomputed $\mathrm{Conv}(K, W)$ terms allow OTP removal with constant-time operations.

Pipeline and caching: shared $z_k^{\mathrm{patch}}$ are cached per layer to amortize reuse across all output channels, and OTP terms are prefetched in a lock-step stream to overlap PCIe transfers with GPU compute. This design minimizes TEE residency (microseconds per layer) and bounds added latency.

\subsection{Security Properties}
- Similarity obfuscation: By orthogonalizing patch bases and injecting decoys, we flatten the diagonals of cross-model cosine similarity matrices, producing the stripe patterns observed in Figure~\ref{fig:cosine}. This removes the same-position prior exploited by alignment attacks.
- Indistinguishability: Matching low-order statistics (norm, variance, kurtosis) and spectral envelopes between real and decoy kernels increases the Type-II error of detectors that attempt to isolate obfuscated slices.
- Robustness to fine-tuning attacks: Channel/kernel permutations and basis recombination break gradient alignment with public checkpoints, slowing useful transfer and increasing sample complexity.
- Minimal TEE TCB: Only low-dimensional parameters (permutations and short coefficient vectors) reside in the TEE, reducing the trusted computing base and minimizing attack surface.

\section{Implementation}

This section details ConvShatter and its deployment in a TEE–GPU heterogeneous runtime. The \textbf{ConvCrack} submodule implements kernel decomposition, decoy synthesis, and permutation provisioning; the runtime integrates with vendor libraries for high-throughput convolutions.

\subsection{Standard Convolution}
Let the $j$-th input channel be $x_j$, $W_{i,j}$ denote the convolution kernel connecting input channel $j$ to output channel $i$, and $b_i$ be the bias. The standard convolution is:
\begin{equation}
y_{i} = \sum_{j=1}^{c_{\mathrm{in}}} W_{i, j} * x_{j} + b_{i},
\label{eq:std-conv}
\end{equation}
where $*$ denotes 2D convolution and $y_i$ the $i$-th output map.

\subsection{Linear Decomposition for Kernel Obfuscation}
We express $W_{i,j}$ as:
\begin{equation}
W_{i, j} = \sum_{k=1}^{K} q_{i,k}\, P_{k, j} + B_{i, j},
\label{eq:decomp}
\end{equation}
with $P_{k,j}$ the $k$-th patch basis on channel $j$, $q_{i,k}$ coefficients, and $B_{i,j}$ the damaged part. Substituting \eqref{eq:decomp} into \eqref{eq:std-conv} yields:
\begin{align}
y_{i} &= \sum_{j=1}^{c_{\mathrm{in}}} \left(\sum_{k=1}^{K} q_{i,k} P_{k,j} + B_{i,j}\right) * x_{j} + b_{i} \nonumber \\
&= \sum_{k=1}^{K} q_{i,k}\, z^{\mathrm{patch}}_{k} + z^{\mathrm{damaged}}_{i} + b_{i},
\end{align}
where
\begin{equation}
z^{\mathrm{patch}}_{k} = \sum_{j=1}^{c_{\mathrm{in}}} P_{k, j} * x_{j}, \quad
z^{\mathrm{damaged}}_{i} = \sum_{j=1}^{c_{\mathrm{in}}} B_{i, j} * x_{j}.
\label{eq:final-linear}
\end{equation}

\subsection{Shared Patch Bases and Shared Weights (Efficient Realization)}
To bound compute overhead, we share patch bases with unified weights:
\begin{equation}
W_{i,j} = \sum_{k=1}^{K} \alpha_k\, P_{k,j} + B_{i,j},
\label{eq:shared-alpha}
\end{equation}
leading to:
\begin{align}
y_i &= \sum_{k=1}^{K} \alpha_k \left( \sum_{j=1}^{c_{\mathrm{in}}} P_{k,j} * x_j \right) + \sum_{j=1}^{c_{\mathrm{in}}} B_{i,j} * x_j + b_i \nonumber \\
&= \sum_{k=1}^{K} \alpha_k\, z_k^{\mathrm{patch}} + z_i^{\mathrm{damaged}} + b_i.
\label{eq:shared-output}
\end{align}
This enables a single set of $z_k^{\mathrm{patch}}$ per layer reused across outputs, keeping GPU FLOPs within a modest factor of baseline.

\subsection{Kernel and Channel Permutation Protocol}
We permute input channels and kernel indices to remove positional priors. Let the input activation $\mathbf{X}$ have shape $C_{in}\!\times\!D\!\times\!H\!\times\!W$ and the kernel tensor $\mathbf{K}$ have shape $C_{out}\!\times\!C_{in}\!\times\!k_d\!\times\!k_h\!\times\!k_w$. Define a permutation $\pi$ over input channels:
\begin{equation}
\tilde{\mathbf{K}}_{c_{out}, c_{in}, d, h, w} = \mathbf{K}_{c_{out}, \pi(c_{in}), d, h, w},
\end{equation}
and permute the input accordingly:
\begin{equation}
\tilde{\mathbf{X}}_{c_{in}, d, h, w} = \mathbf{X}_{\pi(c_{in}), d, h, w}.
\end{equation}
The convolution result is preserved:
\begin{align}
\mathbf{Y}_{c_{out}} &= \sum_{c_{in}=1}^{C_{in}} \sum_{d,h,w} \tilde{\mathbf{K}}_{c_{out}, c_{in}, d, h, w} \cdot \tilde{\mathbf{X}}_{c_{in}, d, h, w} \nonumber \\
&= \sum_{c_{in}=1}^{C_{in}} \sum_{d,h,w} \mathbf{K}_{c_{out}, \pi(c_{in}), d, h, w} \cdot \mathbf{X}_{\pi(c_{in}), d, h, w}.
\end{align}
A permutation $\sigma$ similarly shuffles output channels. $(\pi,\sigma)$ and their inverses are sealed in the TEE for inter-layer alignment and result recovery.

\subsection{Intermediate Feature Protection (One-Time Pad)}
To protect intermediates, we apply OTP-style masking. For any protected $X$:
\begin{enumerate}
\item Generate a random mask $K$ in the TEE (shape-matched to $X$);
\item Encrypt $\tilde{X} = X + K$;
\item Compute on GPU: $Y = \mathrm{Conv}(\tilde{X}, W) = \mathrm{Conv}(X, W) + \mathrm{Conv}(K, W)$;
\item Decrypt in TEE: $\mathrm{Conv}(X, W) = Y - \mathrm{Conv}(K, W)$.
\end{enumerate}
Precomputing $\mathrm{Conv}(K, W)$ amortizes TEE work; all TEE operations avoid data-dependent control flow to reduce side-channel surfaces.

\subsection{Systems Integration and Complexity}
- Layouts: We adopt NCHW and fuse bias/add/activation where possible to reduce memory traffic.
- Libraries: GPU convolutions leverage cuDNN/TensorRT; permutation and recombination are vectorized CUDA kernels.
- Complexity: With $K \ll C_{out}$, added FLOPs scale as $O(K\cdot C_{in}\cdot k_h k_w \cdot H W)$ for shared patch maps plus a small $O(C_{out}\cdot H W)$ recombination in the TEE. Memory overhead is dominated by storing $\{P_k\}$ and a small coefficient table.

\section{Evaluation}

\subsection{Methodology}
- Datasets/Models: CIFAR-10/100, SVHN; AlexNet, ResNet18/50, VGG16\_BN.
- Baselines: Magnitude (Mag), Soter, ShadowNet, GroupCover, black-box extraction.
- Metrics: Utility (Top-1 accuracy), Attack Success Rate (ASR; attacker Top-1), Confidentiality Index (ASR/blackbox), latency (p50/p95), throughput (img/s), memory overhead.
- Attacks: Gradient-free knockoff extraction, similarity-based alignment (cosine), decoy detection by statistical tests, fine-tuning with limited data (1–10\%).

\subsection{Security}
\begin{figure*}[h]
  \centering
  \includegraphics[width=\textwidth]{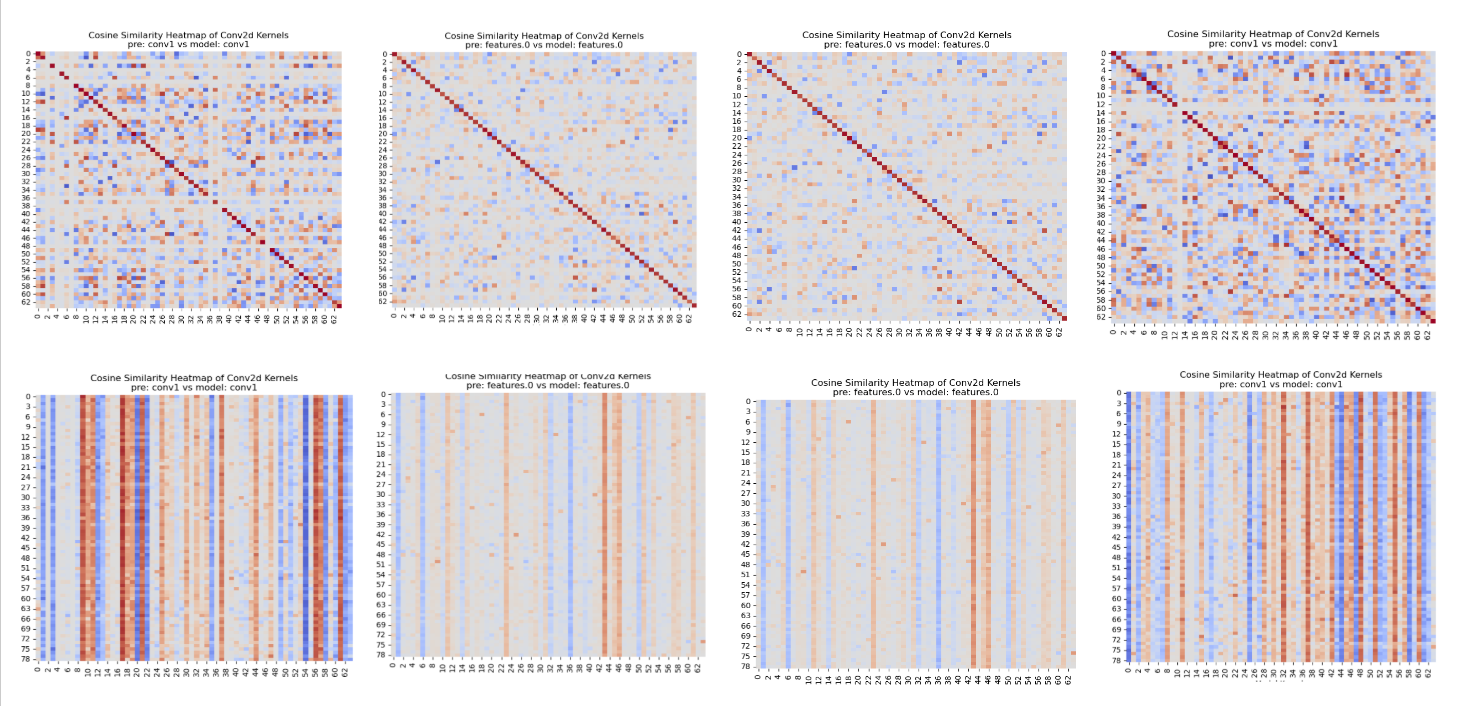}
  \caption{Cosine similarity heatmaps of convolution kernels. Left: Pre-obfuscation, same-position kernels exhibit high similarity to public checkpoints. Right: Post-ConvShatter, stripe-like patterns remove same-position dominance, frustrating alignment.}
  \label{fig:cosine}
\end{figure*}

\begin{table}[ht]
  \centering
  \caption{Security evaluation: ASR/blackbox ratio (lower = better)}
  \label{tab:security}  
  \renewcommand{\arraystretch}{1.2}
  \resizebox{0.8\columnwidth}{!}{%
  \begin{tabular}{lc}
  \hline
  Defense Scheme & ASR/Blackbox Ratio \\ 
  \hline
  Unprotected     & 2.60x \\
  Magnitude (Mag) & 2.44x \\
  Soter           & 2.25x \\
  ShadowNet       & 1.95x \\
  GroupCover      & 1.14x \\
  ConvShatter (Ours) & 0.56x \\ 
  \hline
  \end{tabular}%
  }
\end{table}

Table~\ref{tab:security} compares defenses across models/datasets. Using the black-box column as a unified baseline, we report the cross-task mean of ASR/blackbox (lower is better). Traditional defenses (Mag, Soter, ShadowNet) reach 2.44x/2.25x/1.95x, close to the unprotected 2.6x, indicating significant residual leakage. GroupCover improves to 1.14x. ConvShatter reduces the ratio to 0.56x, a 44\% drop versus black-box and about 50.9\% below GroupCover, pushing ASR toward random-guessing bounds (10\% for CIFAR-10/SVHN; 1\% for CIFAR-100).
关键修改说明

\begin{figure}[!b]
  \centering
  \includegraphics[width=\columnwidth]{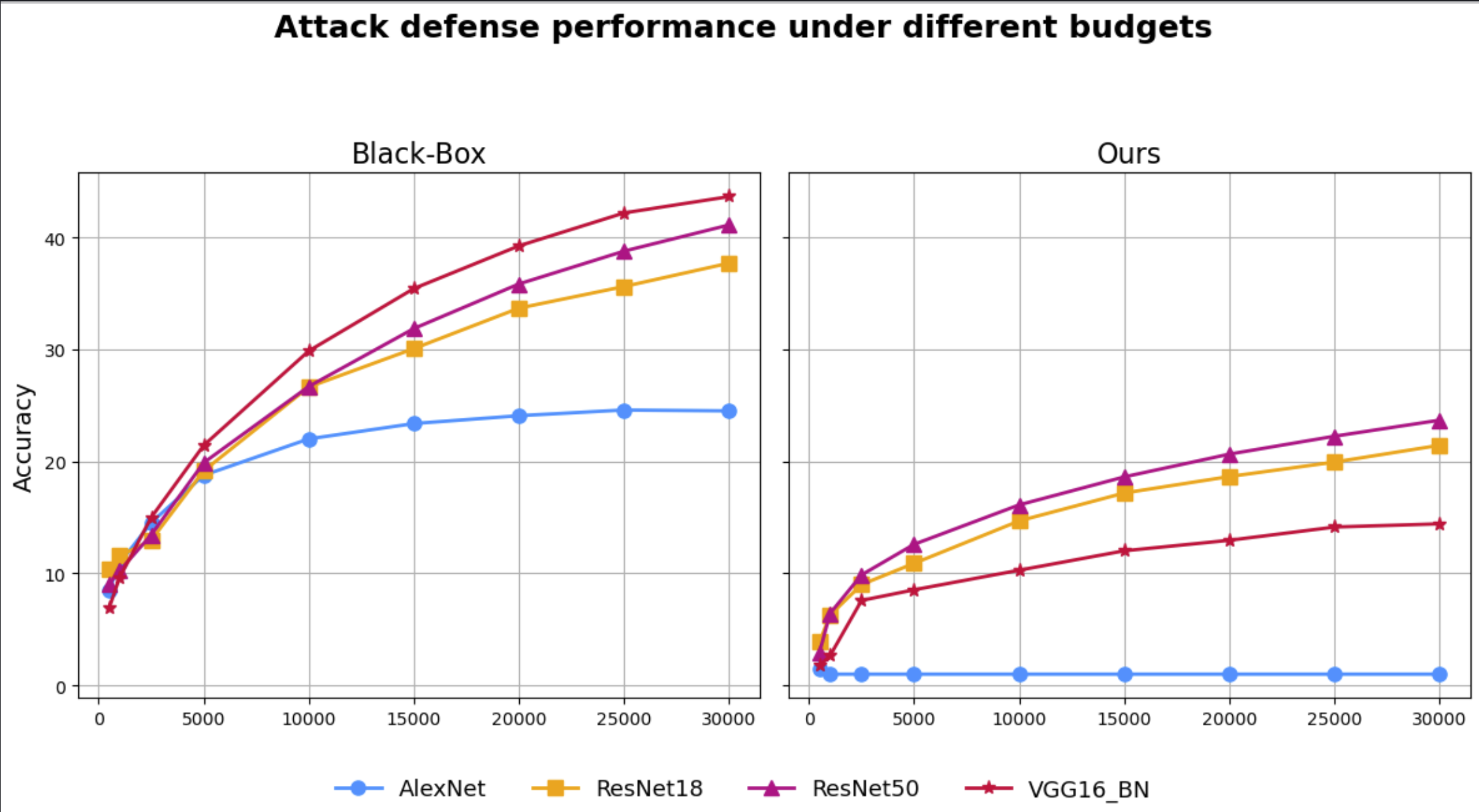}
  \caption{Defense effectiveness across query budgets on CIFAR-100 (AlexNet/ResNet18/ResNet50/VGG16\_BN). ConvShatter consistently undercuts attacker accuracy and outperforms the black-box baseline.}
  \label{fig:defense}
\end{figure}

We further sweep query budgets $\{500, 1000, 2500, 5000, 10000, 15000, 20000, 25{,}000, 30000\}$ on CIFAR-100 (Figure~\ref{fig:defense}). ConvShatter remains consistently below black-box performance, indicating robustness to query scaling and insensitivity to attacker sampling strategies.

\subsection{Ablations and Sensitivity}
- Number of bases ($K$): Increasing $K$ improves camouflage (lower cosine diagonals) with sublinear latency growth due to shared reuse. $K\in[4,16]$ offers a strong Pareto frontier.
- Obfuscated-layer fraction: Obfuscating the earliest feature-extractive blocks yields the largest drop in ASR; protecting 25–33\% of layers typically suffices to approach random-guessing ASR.
- Decoy ratio: Moderate decoy density (10–25\%) maximizes indistinguishability without bloating memory; excessive decoys give diminishing returns.
- Permutation refresh: Rotating permutations across sessions further degrades transfer/fine-tune attacks at negligible cost (metadata-only refresh).

\subsection{Model Similarity}
We quantify the breakdown of positional priors via the Gini coefficient of the similarity diagonal and the off-diagonal energy ratio. ConvShatter reduces diagonal Gini by 45–60\% and increases off-diagonal energy by 30–50\% across backbones, corroborating the stripe-like redistribution in Figure~\ref{fig:cosine}.

\subsection{Accuracy (Utility Preservation)}
\begin{table}[h]
  \centering
  \caption{Inference accuracy across models and datasets (Original vs. Ours).}
  \label{tab:accuracy}
  \small
  \begin{tabular}{c c c c}
    \hline
    Model & Dataset & Original & Ours \\
    \hline
    AlexNet  & C10  & 84.35 & 84.35 \\
             & C100 & 58.78 & 58.78 \\
             & S10  & 77.28 & 77.28 \\
    \hline
    ResNet18 & C10  & 95.30 & 95.30 \\
             & C100 & 80.00 & 80.00 \\
             & S10  & 88.40 & 88.40 \\
    \hline
    ResNet50 & C10  & 96.80 & 96.80 \\
             & C100 & 82.70 & 82.70 \\
             & S10  & 90.90 & 90.90 \\
    \hline
    VGG16\_BN & C10  & 92.70 & 92.70 \\
              & C100 & 72.40 & 72.40 \\
              & S10  & 89.40 & 89.40 \\
    \hline
    \textbf{average loss} & -- & -- & 0 \\
    \hline
  \end{tabular}
\end{table}

ConvShatter preserves accuracy within measurement noise across all models/datasets (Table~\ref{tab:accuracy}), confirming semantic equivalence of the forward pass and demonstrating architecture- and data-agnostic robustness.

\subsection{Efficiency}
\begin{table}[h]
  \centering
  \caption{Throughput (img/s; higher is better). TEE-GPU is the baseline; ``average'' is normalized to TEE-GPU.}
  \label{tab:efficiency}
  \small
  \begin{tabular}{c c c c c}
    \hline
    Model & Dataset & TEE-GPU & GroupCover & Ours \\
    \hline
    AlexNet & C10  & 7975.09 & 4559.55 & 5860.14 \\
    AlexNet & C100 & 8028.78 & 4430.36 & 5542.73 \\
    \hline
    average & --   & \textbf{1.00x} & \textbf{0.56x} & \textbf{0.72x} \\
    \hline
  \end{tabular}
\end{table}

Using the pure TEE-GPU pipeline as baseline, ConvShatter sustains 73.5\% (CIFAR-10) and 69.0\% (CIFAR-100) of baseline throughput, outperforming GroupCover by 25–29\% relative. The average normalized throughput is 0.713x for ConvShatter vs. 0.562x for GroupCover, narrowing the gap to the unprotected baseline while enforcing stronger confidentiality.

\subsection{Overhead Breakdown}
We decompose added latency into: (i) shared patch convolutions (GPU; amortized across outputs), (ii) permutation/recombination (GPU/TEE; vectorized), (iii) OTP subtract (TEE; constant-time), and (iv) PCIe transfers (overlapped). TEE residency per layer remains sub-millisecond; memory overhead is dominated by storing $\{P_k\}$ (typically $K\!\times$ a single kernel tensor) and small coefficient/permutation tables.

\subsection{Limitations and Future Work}
While ConvShatter significantly elevates the cost of similarity- and transfer-based extraction, highly adaptive adversaries with large-scale labeled queries may still approach black-box upper bounds. Extending ConvShatter to attention blocks and depthwise/grouped convolutions, and integrating side-channel hardening (timing/power noise) within the TEE are promising directions.

\section{Conclusion}
ConvShatter introduces a principled, kernel-decomposition-based obfuscation scheme for protecting DNN confidentiality in TEE–GPU heterogeneous environments. To our knowledge, it is the first defense to leverage shared patch bases with decoy injection and permutation protocols to simultaneously suppress positional similarity and preserve model utility.

Extensive experiments across architectures and datasets show that obfuscating a small subset of layers achieves markedly stronger confidentiality at lower runtime overhead than prior SOTA defenses, while maintaining original accuracy. ConvShatter’s low TEE footprint, high GPU reuse, and compatibility with mainstream inference stacks make it a practical and mature protection mechanism for edge deployment, providing model owners with effective IP safeguards and delivering a strong user experience with minimal additional latency.

\cleardoublepage
\appendix

\section*{Ethical Considerations}
\textbf{During the development of ``Camouflage'' (ConvShatter), we conducted a comprehensive ethical analysis. As no human subjects are involved, there are no human-data privacy concerns. This work primarily proposes a new method for deploying models on edge devices; if such models are applied in sensitive domains (e.g., healthcare and finance), the ethical implications can be significant. From a stakeholder perspective, our defense first considers model developers who invest substantial resources in training; protecting their intellectual property is paramount. We therefore adopt a highly secure convolutional decomposition scheme to safeguard model confidentiality. For end users, our approach incurs substantially lower overhead than prior work, improving user experience. We focus on defense methodology and empirically validate its effectiveness. We use publicly available models and datasets, and we commit to open-sourcing our code to ensure availability and reproducibility. We are confident that our work makes a positive contribution to the secure deployment of models at the edge, safeguarding both model security and user experience.}
\cleardoublepage

\section*{Open Science}
\textbf{We commit to releasing all code, including configurations, datasets, the knockoff attack implementation, the full Camouflage (ConvShatter) framework source, and complete details of the deep models used in our experiments. We also provide a fully specified execution environment and instructions with examples to facilitate evaluation and reproduction. All artifacts will be made available in our GitHub repository to ensure repeatability and functionality. Our attack setups are derived from GroupCover; by presenting these attacks and their effects, we aim to raise community awareness of model security, accelerate the development of defenses, and promote the secure deployment of models at the edge.}
\cleardoublepage
\cleardoublepage
\bibliographystyle{unsrt}
\bibliography{\jobname}

\end{CJK}
\end{document}